\documentclass[aps,prb,twocolumn,showpacs,10pt]{revtex4-1}
\usepackage{graphicx}
\usepackage{amssymb}
\usepackage{amsmath}
\usepackage{comment}
\usepackage{footmisc}
\usepackage{color}
\pdfoutput=1

\begin{document}

\def\pc{\frac{2\pi}{\Phi_0}}

\def\e{\varepsilon}
\def\f{\varphi}
\def\p{\partial}
\def\ba{\mathbf{a}}
\def\bA{\mathbf{A}}
\def\bb{\mathbf{b}}
\def\bB{\mathbf{B}}
\def\bD{\mathbf{D}}
\def\bd{\mathbf{d}}
\def\be{\mathbf{e}}
\def\bE{\mathbf{E}}
\def\bH{\mathbf{H}}
\def\bj{\mathbf{j}}
\def\bk{\mathbf{k}}
\def\bK{\mathbf{K}}
\def\bM{\mathbf{M}}
\def\bm{\mathbf{m}}
\def\bn{\mathbf{n}}
\def\bq{\mathbf{q}}
\def\bp{\mathbf{p}}
\def\bP{\mathbf{P}}
\def\br{\mathbf{r}}
\def\bR{\mathbf{R}}
\def\bS{\mathbf{S}}
\def\bu{\mathbf{u}}
\def\bv{\mathbf{v}}
\def\bV{\mathbf{V}}
\def\bw{\mathbf{w}}
\def\bx{\mathbf{x}}
\def\by{\mathbf{y}}
\def\bz{\mathbf{z}}
\def\bG{\mathbf{G}}
\def\bW{\mathbf{W}}
\def\Bn{\boldsymbol{\nabla}}
\def\Bo{\boldsymbol{\omega}}
\def\Br{\boldsymbol{\rho}}
\def\Bs{\boldsymbol{\hat{\sigma}}}
\def\bh{{\beta\hbar}}
\def\mA{\mathcal{A}}
\def\mB{\mathcal{B}}
\def\mD{\mathcal{D}}
\def\mF{\mathcal{F}}
\def\mG{\mathcal{G}}
\def\mH{\mathcal{H}}
\def\mI{\mathcal{I}}
\def\mL{\mathcal{L}}
\def\mO{\mathcal{O}}
\def\mP{\mathcal{P}}
\def\mT{\mathcal{T}}
\def\mU{\mathcal{U}}
\def\mZ{\mathcal{Z}}
\def\fr{\mathfrak{r}}
\def\ft{\mathfrak{t}}
\newcommand{\rf}[1]{(\ref{#1})}
\newcommand{\al}[1]{\begin{aligned}#1\end{aligned}}
\newcommand{\ar}[2]{\begin{array}{#1}#2\end{array}}
\newcommand{\eq}[1]{\begin{equation}#1\end{equation}}
\newcommand{\bra}[1]{\langle{#1}|}
\newcommand{\ket}[1]{|{#1}\rangle}
\newcommand{\av}[1]{\langle{#1}\rangle}
\newcommand{\AV}[1]{\left\langle{#1}\right\rangle}
\newcommand{\aav}[1]{\langle\langle{#1}\rangle\rangle}
\newcommand{\braket}[2]{\langle{#1}|{#2}\rangle}
\newcommand{\ff}[4]{\parbox{#1mm}{\begin{center}\begin{fmfgraph*}(#2,#3)#4\end{fmfgraph*}\end{center}}}

\def\mr{m_{\perp}}
\def\ml{m_{\parallel}}
\def\hr{H_{\perp}}
\def\hl{H_{\parallel}}

\def\mb{(\mu+\alpha\nu)}
\def\nb{(\nu-\alpha\mu)}
\def\lb{(\lambda+\alpha\kappa)}
\def\kb{(\kappa-\alpha\lambda)}
\def\mn{\left|\bm\times\bz\right|}
\def\etap{\frac{2\pi}{\Phi_0}}
\def\ab{\bar{\alpha}}

\title{Impurity Induced Quantum Phase Transitions and Magnetic Order in Conventional Superconductors: Competition between Bound and Quasiparticle states}

\author{Silas Hoffman$^1$}
\author{Jelena Klinovaja$^1$}
\author{Tobias Meng$^{2,1}$}
\author{Daniel Loss$^1$}
\affiliation{$^1$Department of Physics, University of Basel, Klingelbergstrasse 82, CH-4056 Basel, Switzerland}
\affiliation{$^2$Institut f\"ur Theoretische Physik, Technische Universit\"at Dresden, 01062 Dresden, Germany}

\begin{abstract}
We theoretically study bound states generated by magnetic impurities within conventional $s$-wave superconductors, both analytically and numerically. In determining the effect of the hybridization of two such bound states on the energy spectrum as a function of magnetic exchange coupling, relative angle of magnetization, and distance between impurities, we find that quantum phase transitions can be modulated by each of these parameters. Accompanying such transitions, there is a change in the preferred spin configuration of the impurities. Although the interaction between the impurity spins is overwhelmingly dominated by the quasiparticle contribution, the ground state of the system is determined by the bound state energies. 
Self-consistently calculating the superconducting order parameter, we find a discontinuity when the system undergoes a quantum phase transition as indicated by the bound state energies.
\end{abstract}


\maketitle

\textit{Introduction}.---In a conventional $s$-wave superconductor, quasiparticle excitation energies are separated from the chemical potential due to the formation of the superconducting gap. When magnetic impurities are present, the exchange interaction can induce a bound state within the gap known as a Yu-Shiba-Rusinov (YSR) state, \cite{yuAPS65,*shibaPTP68,*rusinovJETP69} which has been studied in detail both experimentally and theoretically. \cite{sakuraiPoTP70,yazdaniSCI97,flattePRL97,flattePRB97,salkolaPRB97,flatteSSP99,
flattePRB00,balatskyRMP06,zhangPRL08,yaoPRL14,yaoPRB14,zyuzinPRB14,mengCM15,Franke} Recently, these states have attracted much attention in the context of magnetic impurity chains in which, when sufficiently close together, individual YSR states can hybridize with adjacent bound states forming a band within the superconducting gap that can host Majorana fermions at its ends. \cite{nadj-pergePRB13,klinovajaPRL13,vazifehPRL13,brauneckerPRL13,nakosaiPRB13,
pientkaPRB13,poyhonenPRB14,reisPRB14,nadj-pergeSCI14,sau,schon}

Two magnetic impurities interacting via quasiparticles are well described by the Ruderman-Kittel-Kasuya-Yosida (RKKY) interaction \cite{rudermanPR54,*kasuyaPTP56,*yosidaPR57,simon2007:PRL,braunecker2009:PRL,Braunecker_Jap_Klin_2009,QHE} when the exchange interaction between the impurity and quasiparticles is much smaller than the Fermi energy. This results in a noncollinear orientation between the impurities in three-dimensional superconductors. \cite{andersonPR59,abrikosovBK88} Although for many parameters the contribution to the inter-impurity exchange mediated by the overlap of YSR states is much smaller than that of the quasiparticles, \cite{abrikosovBK88,aristovZPB97,galitskiPRB02} it has been shown that resonant YSR bound states can dominate the exchange interaction and induce an antiferromagnetic alignment of the impurities. \cite{yaoPRL14} However, for the experimentally relevant limit \cite{nadj-pergeSCI14} when the exchange interaction is of the order the Fermi energy, a theoretical understanding of the interaction between magnetic impurities including (1) the quasiparticle contribution and (2) a self-consistent local reduction of the gap is missing from the literature. 

In this Letter, we determine the interaction between two  magnetic impurities for arbitrary angles between their spins wherein the strength of the exchange interaction is unrestricted and, in general, unequal at the sites of the impurities. First, by analytically calculating the bound state energy spectrum, we find that a quantum phase transition (QPT)~\cite{flattePRL97,flattePRB97,salkolaPRB97,flatteSSP99,flattePRB00,balatskyRMP06,mengCM15} can be tuned by changing the distance between and relative magnetic orientation of the impurities. We, numerically, include the bulk contribution to the exchange interaction which quantitatively dominates over the YSR contribution for many parameters. \cite{abrikosovBK88,aristovZPB97,galitskiPRB02,yaoPRL14} Further, carrying out self-consistently calculations, we find a discontinuity in the superconducting order parameter when the system undergoes such a QPT as indicated by the bound state energies. This, in turn, gives rise to magnetic metastable states, in addition to the lowest energy magnetic configuration, for a sufficiently large exchange interaction.

\begin{figure}[pt]
\includegraphics[width=0.7\linewidth,clip=]{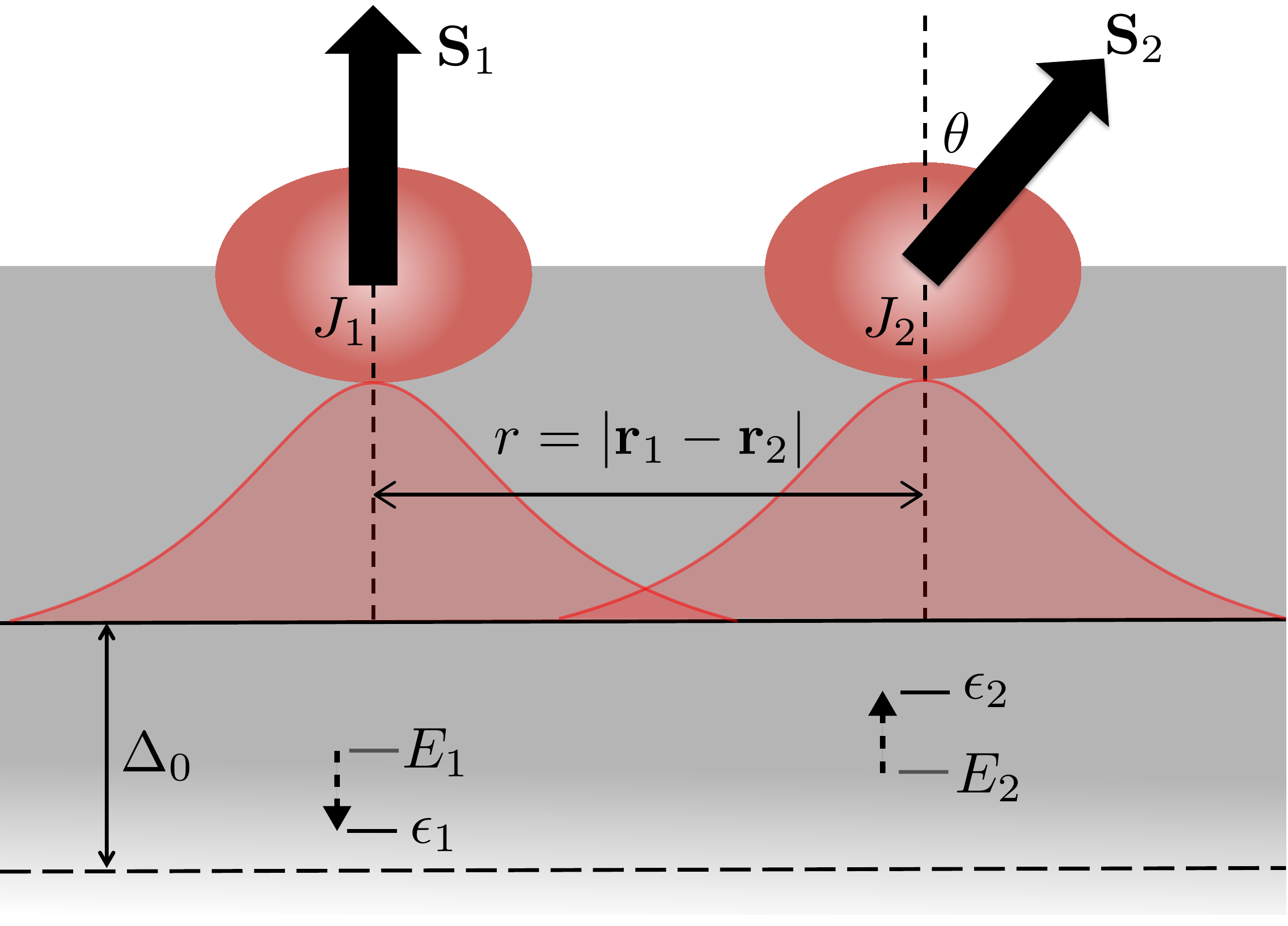}
\caption{Our setup of two magnetic impurities at $\textbf{r}_1$ and $\textbf{r}_2$ in an $s$-wave superconductor with classical spins $\textbf{S}_1$ and $\textbf{S}_2$, respectively, oriented at a relative angle $\theta$. As a result of the magnetic exchange couplings, $J_1$ and $J_2$, YSR bound states form within the bulk gap $\Delta_0$.  When the distance between the impurities, $r$, is larger than the coherence length of the superconductor the energies are $E_1$ and $E_2$ but get changed to $\epsilon_1$ and $\epsilon_2$ as $r$ decreases and the bound states hybridize with each other.}
\label{device}
\end{figure}

\textit{Model}. We consider two magnetic impurities embedded in a bulk $s$-wave superconductor, see Fig. \ref{device}.  The quasiparticles interact with the impurity spin through the exchange interaction that produces a local effective magnetic field. The corresponding Bogoliubov-de Gennes Hamiltonian density is given by 
\eq{
 H=\xi_\textbf{p}\tau_z+\Delta(\textbf r)\tau_x -\sum_{i=1,2} J_i\textbf{S}_i\cdot\boldsymbol\sigma\delta(\textbf{r}_i-\textbf{r})\,,
\label{BdG}
}
where $\xi_\textbf{p}$ is the dispersion of the quasiparticles with momentum $\bf p$ in the normal metal phase and $\Delta(\textbf r)$ is the local superconducting pairing strength. The Pauli matrices $\boldsymbol \tau$ ($\boldsymbol\sigma$) act in Nambu (spin) space. 
The exchange coupling strength $J_i$  of the spin impurity at $\textbf{r}_i$  can be positive or negative  corresponding to ferro- or antiferromagnetic interactions with quasiparticles, respectively. Here, we focus on $J_{i}>0$ without loss of generality. We assume that $\textbf S_i$ are the classical spin vectors of the impurity at $\textbf{r}_i$, and $\theta$ is the angle between them. The magnitude of the spins, $S_i=\left|\textbf S_i\right|$, are much larger than $\hbar$  so that quantum mechanical spin fluctuations, e.g. the Kondo effect, are negligible. In the following analytics we assume that  $\Delta(\textbf r)=\Delta_0$ is spatially uniform and neglect its suppression due to the impurities, \cite{salkolaPRB97, rusinovJETP69b,*schlottmanPRB76}  which we account for  self-consistently in the numerics following earlier work.\cite{flattePRL97,flattePRB97,salkolaPRB97,flatteSSP99,flattePRB00,balatskyRMP06,mengCM15}

To determine the energy of the bound states, $\epsilon_{1,2}$ and $-\epsilon_{1,2}$ (particle-hole symmetry), we apply a straightforward calculation along the lines of Ref. \onlinecite{pientkaPRB13} and obtain a coupled set of secular equations for the BdG four-component spinors $\psi(\textbf{r})$ at $\textbf{r}_1$ and $\textbf{r}_2$, 
\begin{align}
\psi(\textbf r_1)&=\hat J_1 \textbf s_1\cdot\boldsymbol\sigma\psi(\textbf r_1)+\hat \Gamma_2 \textbf s_2\cdot\boldsymbol\sigma\psi(\textbf r_2)\,,\nonumber\\
\psi(\textbf r_2)&=\hat J_2 \textbf s_2\cdot\boldsymbol\sigma\psi(\textbf r_2)+\hat\Gamma_1 \textbf s_1\cdot\boldsymbol\sigma\psi(\textbf r_1)\,,
\label{EoM2}
\end{align}
where $\textbf s_i=\textbf S_i/S_i$ and
\begin{align}
&\hat J_i=\frac{\alpha_i \left(
\epsilon +\tau_x \Delta_0
\right)}{\sqrt{\Delta_0^2-\epsilon^2}},\\
&\hat\Gamma_i=\alpha_i\Big(\frac{ (\epsilon+\tau_x \Delta_0  )\sin k_F r} {\sqrt{\Delta_0^2-\epsilon^2}}+\tau_z \Big)
\frac{e^{-r/\xi_\epsilon}}{k_F r}
\end{align}
for $i=1,2$. Here, $\alpha_i=\nu_0\pi J_i S_i$, where $\nu_0$ is the density of states evaluated at the Fermi energy, $r=|\textbf r_1-\textbf r_2|$ is the distance between the impurities, 
$k_F$ ($v_F$) is the Fermi wave vector (Fermi velocity) and  $\xi_\epsilon=v_F/\sqrt{\Delta_0^2-\epsilon^2}$.  When the distance between impurities is much greater than the superconducting coherence length, $r\gg\xi_0$, the impurities effectively decouple, $\hat\Gamma_i\rightarrow0$, and one finds that Eq.~(\ref{EoM2}) furnishes solutions that are non-overlapping  YSR bound states at $\textbf{r}_1$ and $\textbf{r}_2$ with energies   $\pm E_i=\pm\Delta_0(1-\alpha_i^2)/(1+\alpha_i^2)$.\cite{yuAPS65,*shibaPTP68,*rusinovJETP69,pientkaPRB13}  In this limit, for sufficiently large exchange interaction, $J_i>1/\nu_0\pi S_i$, the bound state energy goes below the chemical potential and the system undergoes a QPT wherein the parity of the ground changes.\cite{flattePRL97,flattePRB97,salkolaPRB97,morrPRB06}

In order to determine the energies of the hybridized bound states analytically from Eq. (\ref{EoM2}), we focus on distances between impurities much smaller than the coherence length, $r\ll\xi_0$, so that $e^{-r/\xi_\epsilon}\approx 1$ and the hybridization is determined to leading order by $1/k_Fr$. Formally diagonalizing the Hamiltonian and using a variational wave function as an ansatz for the ground state,\cite{Note1} the total energy of the system $\mathcal{E}_{gr}$ is given by \cite{salkolaPRB97,yaoPRL14}
 \eq{
\mathcal{E}_{gr}(\theta)=-\frac{1}{2}\sum_n |\epsilon_n(\theta)|,
\label{totalE}
} 
where $n$, in general, runs over all solutions to Eq.~(\ref{BdG}); in the following analytics we only sum the bound state energies and determine $\mathcal{E}(\theta) =-( |\epsilon_1(\theta)|+|\epsilon_2(\theta)|)/2$.

\textit{Weak hybridization.} For the moment, we consider the case of weak hybridization ($k_Fr \gg E_i/\Delta_0$) for YSR states sufficiently far away from the chemical potential, so that the occupation of the bound states, and thus the ground state, is fixed by $\alpha_i$. That is, when $\alpha_i<1$ ($\alpha_i>1$) the energy is above (below) the chemical potential. Calculating the full analytic solution and then expanding to second order in $1/k_Fr$, which is valid when $|1-\alpha_i|k_F r\gg 1$ and $|\alpha_1-\alpha_2|k_F r\gg 1$, \cite{Note2} the spectrum has two solutions of the form
\eq{
\epsilon_n(\theta)\approx E_n +\Delta_0( A_n +B_n \cos\theta)/(k_F r)^2,
\label{Eweak}
}
where the coefficients  $A_n$ and $B_n$  are functions of $\alpha_1$, $\alpha_2$, and $k_F r.$ \cite{Note1} The bound state energy is extremized when either $\theta=0$ or $\pi$, i.e. the groundstate of impurities is collinear.
When $\epsilon_1\epsilon_2>0$,  $\mathcal E(\pi)$ is always smaller than $\mathcal E(0)$ \cite{yaoPRL14,Note1} and therefore the ground state is antiferromagnetic. When $\epsilon_1 \epsilon_2<0$, $\mathcal E(\pi)>\mathcal E(0)$ and a ferromagnetic orientation is favored.

\textit{Strong hybridization: identical impurities}. Although strong hybridization between impurities cannot be addressed perturbatively, in the symmetric case of equal exchange coupling, {\it i.e.} $\alpha_1=\alpha_2\equiv \alpha$, 
Eq. (\ref{EoM2}) can be solved directly. Because 
the analytic solution for arbitrary $\theta$ is too involved, we focus here on collinear alignments. In the ferromagnetic configuration, the bound state energy levels are given by
\eq{
\epsilon^F_\pm\equiv\epsilon_{1,2}(0)=-\Delta_0(a\pm b)/\sqrt{(a\pm b)^2+c^2}\,,
\label{Eferr}
}
where $a$, $b$, and $c$, which depend on $k_F$, $r$ and $\alpha$ are discussed in Ref.~\onlinecite{Note1}.
The initially twofold degenerate energy levels of the bound states are both split due to hybridization and shifted due to the effective Zeeman splitting at both $\textbf{r}_1$ and $\textbf{r}_2$.

In the antiferromagnetic configuration, the energy level stays twofold degenerate \cite{flattePRB00} and is given by  \cite{Note1} 
\begin{align}
\epsilon^A\equiv\epsilon_{1,2}(\pi)&=\Delta_0\sqrt{\frac{(1-\alpha^2)^2+2({\alpha}/{k_F r})^2+d}{(1+\alpha^2)^2+2 ( {\alpha}/{k_F r})^2 
 \cos 2k_F r+d}}\,.
\label{Eanti}
\end{align} 

\begin{figure}[pt]
\includegraphics[width=0.9\linewidth]{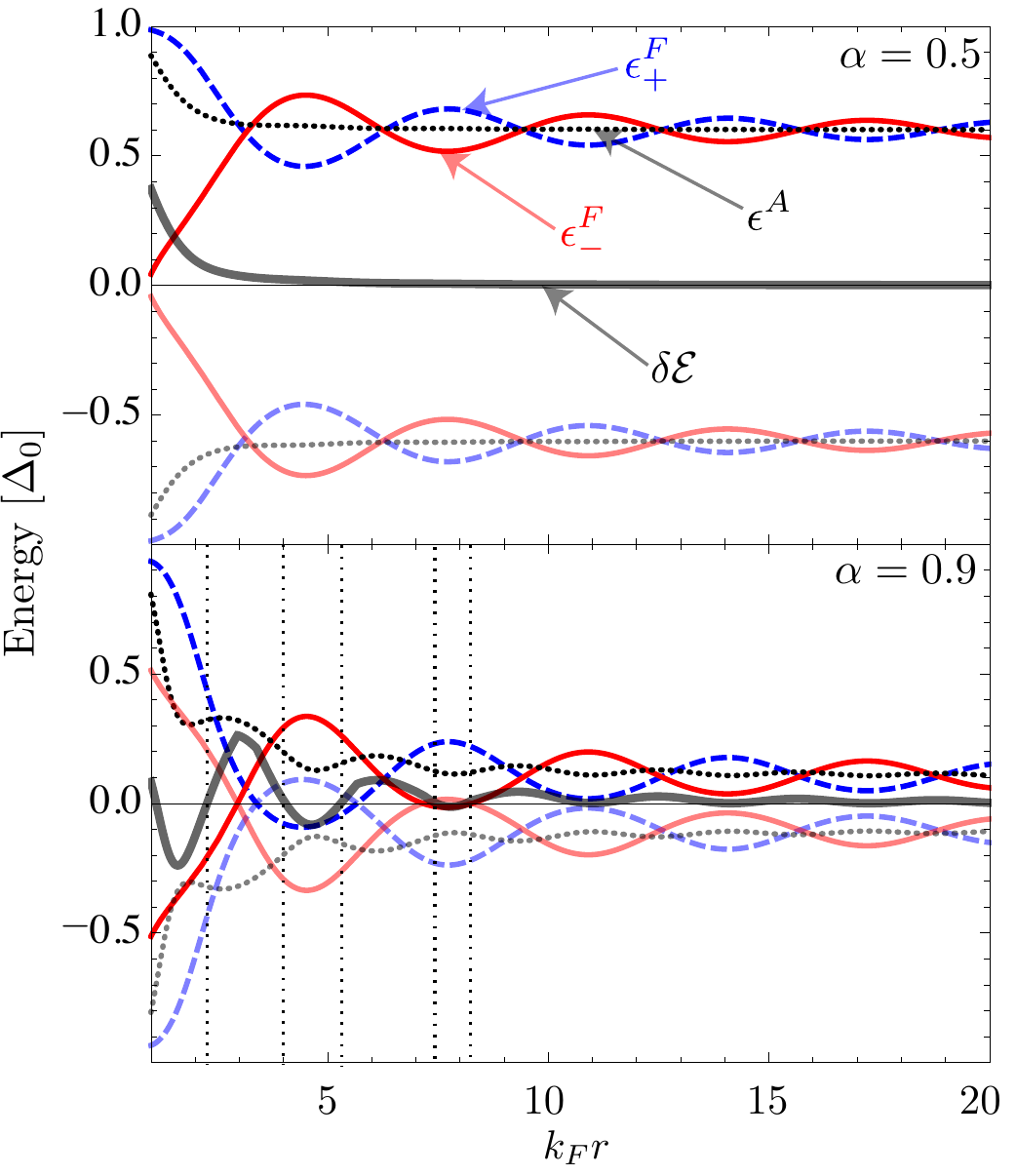}
\caption{ 
Energy of the YSR bound states for the identical magnetic impurities  oriented ferromagnetically (solid and dashed) and antiferromagnetically (dotted) as well as the energy difference $\delta\mathcal{E}$ (thick solid lines) as a function of the distance $r$ between impurities. When $\alpha=0.5$ (top panel), the system remains antiferromagnetic ($\delta\mathcal{E}>0$), while for $\alpha=0.9$ (lower panel) the magnetic configuration oscillates between being antiferromagnetic and ferromagnetic. For convenience, these two configurations are separated by the vertical dotted lines. }
\label{bTot}
\end{figure}

The difference in YSR bound state energy between the two collinear configurations, $\delta \mathcal{E}\equiv \mathcal{E}(0)-\mathcal{E}(\pi)=-(|\epsilon^+_F|+|\epsilon^-_F|-2|\epsilon^A|)/2$, as a function of $k_Fr$ is shown in Fig. \ref{bTot}.  When $\alpha=0.5$ [Fig.~\ref{bTot} (upper panel)], all the electron-like energies in either configuration are greater than zero, in the displayed range, $k_Fr\geq 1$. Furthermore, $\delta \mathcal{E}>0$ and therefore the exchange interaction between impurities is antiferromagnetic, in agreement with the weak coupling limit. If the impurity levels are close to the chemical potential, e.g. $\alpha=0.9$ [Fig.~\ref{bTot} (lower panel)], the ground state of the system depends on the distance between the impurities. When $r$ is sufficiently large, so that the condition for weak hybridization is met, the preferred ordering is antiferromagnetic. When $k_Fr\approx8$, $\epsilon_F^-$ goes below the chemical potential. Near this value of $k_Fr$, $\delta \mathcal{E}$ becomes negative and therefore the preferred magnetic ground state is ferromagnetic rather than antiferromagnetic. As the distance between the impurities decreases further, the bound state energies oscillate about the chemical potential as a function of $r$, thereby changing the YSR ground state. As a result, $\delta\mathcal{E}$ also oscillates around zero implying a change between ferromagnetic and antiferromagnetic configurations.

\begin{figure}[pt]
\includegraphics[width=0.9\linewidth]{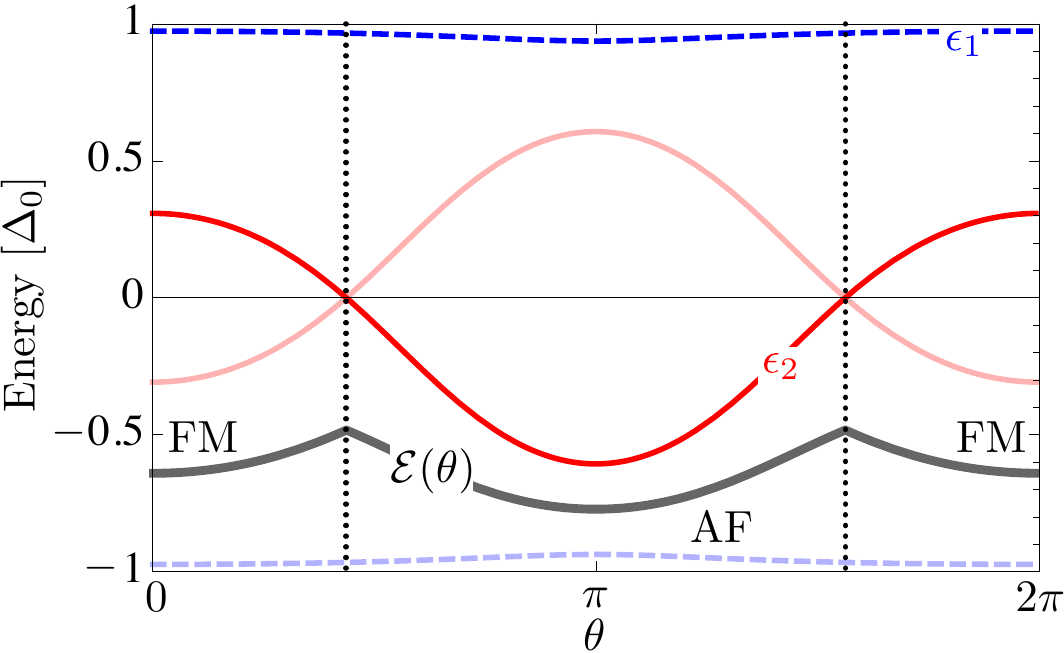}
\caption{ The energy of the bound states ($\epsilon_1$,  $\epsilon_2$) and the total YSR state energy ($\mathcal{E}(\theta)$) as a function of relative angle $\theta$ for $k_Fr=1$, $\alpha_1=0.5$, and $\alpha_2=1$. The change of quantum ground state at $\theta\approx\pm\pi/2$ is indicated by vertical dotted lines.
}
\label{arbT}
\end{figure}

\textit{Angle controlled quantum phase transition.} As seen in the previous section, for some values of $r$ ($|E_i|/\Delta_0\lesssim k_Fr$), the bound state energies are on opposite sides of the chemical potential in the ferromagnetic configuration due to hybridization, while in the antiferromagnetic the energies are always degenerate. Therefore, quite remarkably, one may drive a QPT by changing the relative angle of the impurities. 
As shown  in Fig.~\ref{arbT}, one of YSR bound states passes through the chemical potential at $\theta\approx\pi/2$, signaling a QPT. 
The energy of YSR states is a minimum for the antiferromagnetic configuration, $\theta=\pi$. Decreasing the angle between the impurities increases the energy until a critical point ($\theta\approx\pm\pi/2$) when the ferromagnetic configuration becomes a minimum. Therefore, while the parameters chosen favor an antiferromagnetic configuration as an \textit{absolute} ground state, they additionally support a \textit{metastable} ferromagnetic configuration.\cite{Note3}

\textit{Bulk contribution}.
To address the contributions coming from the bulk, we follow earlier work~\cite{flattePRL97,flattePRB97,salkolaPRB97,flatteSSP99,flattePRB00,balatskyRMP06,mengCM15} and study numerically a two-dimensional system with two magnetic impurities determining 
self-consistently the renormalization of the gap which cannot be addressed analytically.\cite{rusinovJETP69b,*schlottmanPRB76,mengCM15} We use the tight-binding Hamiltonian 
\begin{align}
&\bar H= -t \sum_{<i,i'>}\sum_{\sigma=\pm1} c_{i\sigma}^\dagger c_{i'\sigma}  + \sum_{i} ( \Delta_i  c_{i  1} c_{i \bar 1} + { H.c.}) 
\nonumber \\
&+ \sum_{i} \sum_{\sigma=\pm 1} \big( [\mu - 4t +(\delta_{i1}+\delta_{i2})\bar J_i \sigma \cos \theta_i]c_{i\sigma}^\dagger c_{i\sigma} 
\nonumber \\
 &\hspace{80pt}+ (\delta_{i1} + \delta_{i2})\bar J_i \sin \theta_i \,c_{i\sigma}^\dagger c_{i\bar \sigma} \big),
\label{tbH}
\end{align}
where $c_{i\sigma}$ is the annihilation operator acting on an electron with spin $\sigma$ at lattice site $i$, and the first sum runs over neighboring sites $i$ and $i'$ located in a two-dimensional square lattice of size $N_x \times N_y$ with lattice constant $a$. The chemical potential $\mu$ is taken from the bottom of the energy band, and the local order parameter  $\Delta_i$ is determined self-consistently in an iterative fashion for fixed values of the exchange coupling $\bar J_i$ at site $i$ starting from the uniform superconducting order parameter $\Delta_0$. To compare to the analytics, we consider two impurities located at $i=1$ and $i=2$ (which are not necessarily adjacent) with equal exchange coupling, $\bar J=\bar J_1=\bar J_2$, and fixing the difference in magnetic orientation to be $\theta$, mirroring the schematics of Fig.~\ref{device}. After numerically diagonalizing Eq.~(\ref{tbH}), we find two types of energies in the spectrum: the energy of two YSR bound states  $\mathcal{E} (\theta)$ considered before analytically and the total bulk energy $\mathcal{E}_{gr}(\theta)$ obtained by summing all the energies below the chemical potential, see Eq.~(\ref{totalE}).

\begin{figure}[pt]
\includegraphics[width=0.9\linewidth]{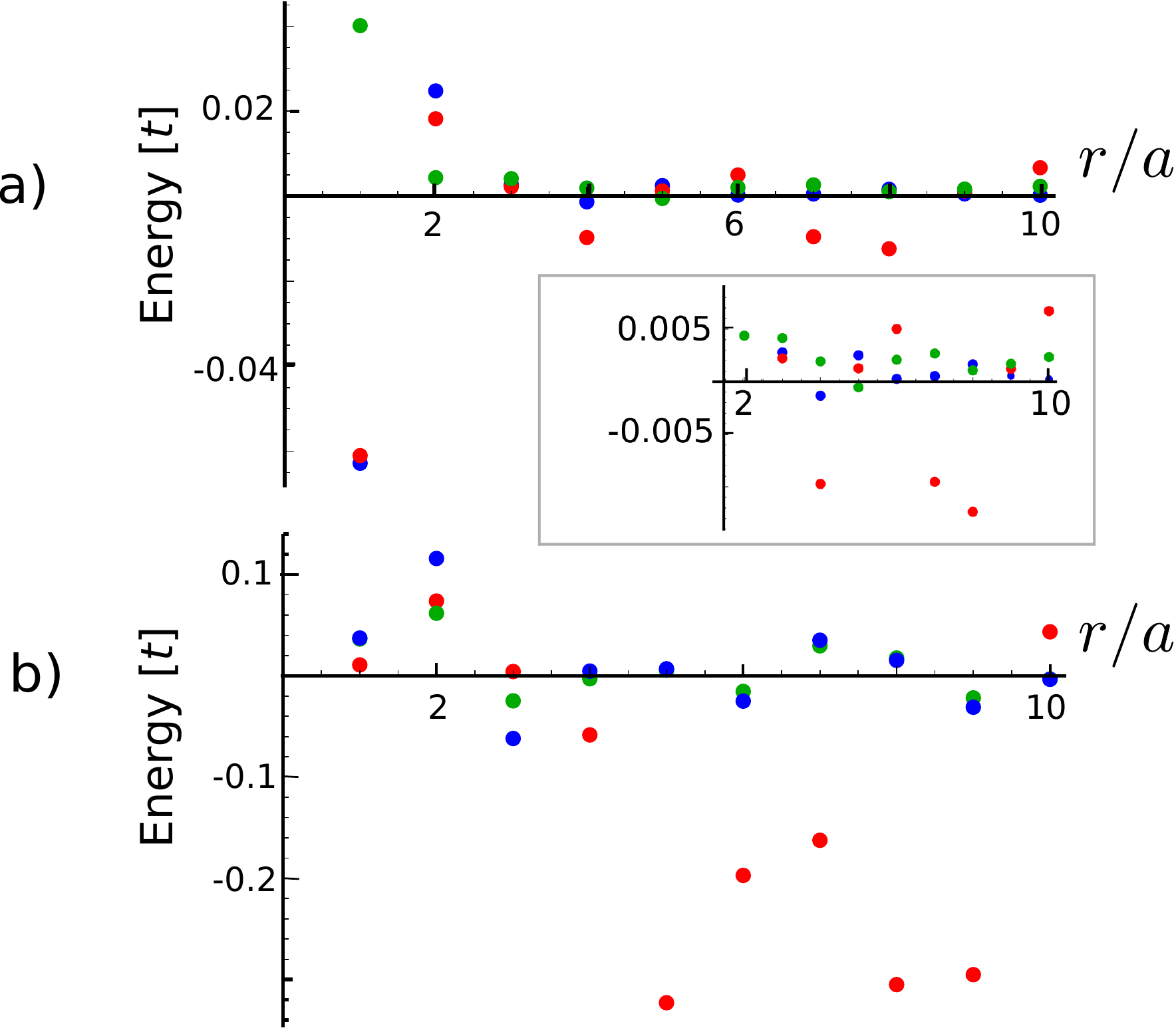}
\caption{The difference in the energy $\delta \mathcal{E}_{gr}$ between ferro- and antiferromagnetic configurations of the system consisting of two identical impurities of coupling strength $\bar J$   as a function of the distance between impurities, $r/a$ for (a)  $\bar J/t=1$ and (b)  $\bar J/t=2.5$ found  self-consistently (red dots) and not self-consistently (blue dots).  Insert: enlarged area of (a) for large distances. The difference in the energy  $\delta\mathcal{E}$ between ferro- and antiferromagnetic configurations including only the YSR bound state (green dots) is found self-consistently. The parameters used are $N_x\times N_y=33\times 25$, $\mu/t=1$, and $\Delta_0/t=0.1$.
 }
\label{bulk}
\end{figure}

First,  we consider the difference between the ground state energies in the collinear magnetic configurations, $\delta\mathcal{E}$ and  $\delta \mathcal{E}_{gr}$  as a function of distance $r$, see Fig.~\ref{bulk}.
The YSR bound state contribution $\delta\mathcal{E}$  is positive for nearly all values of $r$, when $\bar J/t=1$ [Fig.~\ref{bulk}(a)],  {\it i.e.} antiferromagnetic configuration is preferred.  Whereas for $\bar J/t=2.5$ [Fig.~\ref{bulk}(b)], $\delta\mathcal{E}$  oscillates between positive and negative values. Both results agree with the analytics.

Second, we aim to address the effect of gap renormalization and plot  $\delta \mathcal{E}_{gr}$  without self-consistent renormalization assuming $\Delta_i\equiv\Delta_0$ (see  Fig.~\ref{bulk}).
Interestingly,  $\delta \mathcal{E}_{gr}$ is changed only slightly for all values of $\bar J$, keeping the energies at the same order of magnitude. Upon including renormalization of the gap, $\delta \mathcal{E}_{gr}$  is increased drastically and the magnetic orientation becomes very sensitive to the distance between the impurities. This emphasizes the importance of a self-consistent renormalization of the gap when calculating the energies of such a system especially close to the phase transition.

\begin{figure}[!th]
\includegraphics[width=0.7\linewidth]{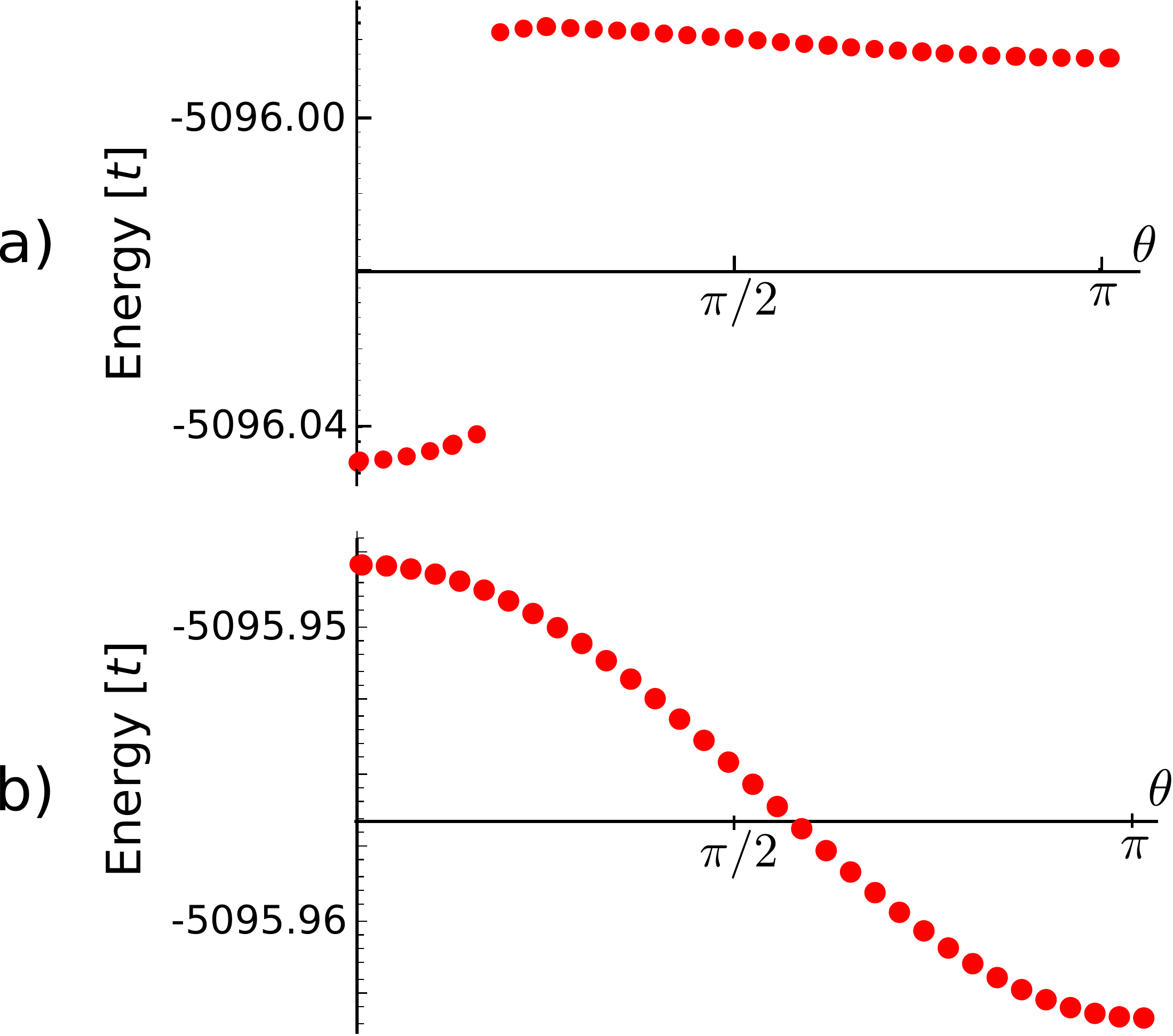}
\caption{The total energy of the ground state for two identical impurities ($\bar J=\bar J_1=\bar J_2$) as a function of the angle $\theta$ between magnetic moments  found numerically for (a) $\bar J/t=2.17$ (ferromagnetic ordering) and (b) $\bar J/t=2$ (antiferromagnetic ordering) at the distance  $r/a=6$. Other parameters are the same as in Fig. \ref{bulk}.
}
\label{FullEnergy}
\end{figure}

Third, we determine the angular dependence of the total energy and YSR bound state energy.
For the fixed distance between the impurities (see Fig. \ref{FullEnergy}), we observe that away from the phase transition,  $\mathcal{E}_{gr}$ and $\mathcal{E}$ changes monotonically for $\theta \in [0,\pi]$, and, thus, the ground state is either ferromagnetic or antiferromagnetic which is consistent with analytical results. In contrast to that, close to the phase transition when the bound state energies do cross the chemical potential as a function of $\theta$, the dependence is non-monotonic [see Fig. \ref{FullEnergy}(a)], and, in addition to the ferromagnetic (antiferromagnetic) ground state, there is a metastable antiferromagnetic (ferromagnetic) state. We also note that self-consistent solution  demonstrates a jump in energy as one of YSR states crosses zero energy.
Thus, we again find the qualitative agreement with analytical calculations predicting metastable states by analyzing only YSR bound states.  However, we emphasize that it is the QPT that results in the metastable state in Fig.~\ref{FullEnergy}(a), while the interaction is dominated by the bulk (not bound) state contribution to the energy.

\textit{Conclusions}.---We have studied how the orientation of two spin impurity  coupled via overlap of the YSR bound states induced by them depends on the distance between impurities and the strength of the exchange interaction. We have also demonstrated that a QPT
can be controlled by changing relative magnetic orientation. Generally, the bulk contribution to the total ground state energy dominates over the bound state contribution, especially if the superconducting order parameter is determined self-consistently. The proposed effects could be measured with STM \cite{STM_review} or NV-center \cite{NV,RKKY_measure} techniques. 

\begin{acknowledgments}
We acknowledge support from the Swiss NSF, NCCR QSIT, and the DFG through GRK 1621 and SFB 1143. 
\end{acknowledgments}


%

\appendix

\begin{widetext}

\section{Variational Wave Function}
\label{variational}
We extending the variational wave function for one impurity \cite{salkolaPRB97} to two impurities. For sufficiently weak coupling, in both the exchange interaction ($\alpha_1$, $\alpha_2\lesssim1$) and the bound state hybridization ($k_Fr\gtrsim1$), the ground state is given by the BCS-like wave function $|\Psi_0\rangle\sim\prod_{n>0}(u_n+v_n\psi_{n}^\dagger\psi_{-n}^\dagger|0\rangle$,
where $\psi_n$ furnish a basis for the BdG Hamiltonian in the presence of the impurities for a given magnetic alignment and $u_n$ and $v_n$ are the Bogoliubuv coherence factors. The quasiparticle operators $\gamma_n$ are defined as $\gamma_1=u_1\psi_1-v_1\psi_{-1}^\dagger$, $\gamma^\dagger_{-1}=u_1\psi_{-1}+v_1\psi_{1}^\dagger$, $\gamma^\dagger_1=u_1\psi^\dagger_1-v_1\psi_{-1}$, and similarly for $n>1,$ so that $\gamma_n|\Psi_0\rangle=0$ for all $n$. Let $n=1$ correspond to the lower energy bound state and $n=2$ to the higher one while $-n$ corresponds to a state with reversed spin.  When the lower energy bound state is occupied, the wavefunction is given by $|\Psi_1\rangle\sim\gamma_1^\dagger|\Psi_0\rangle=\psi_1^\dagger\prod_{n>1}(u_n+v_n\psi_n^\dagger\psi_{-n}^\dagger)|0\rangle$. When both states are occupied, the wavefunction is $|\Psi_{1,2}\rangle\sim\gamma_2^\dagger|\Psi_1\rangle=\gamma_2^\dagger\gamma_1^\dagger|\Psi_0\rangle=\psi_2^\dagger\psi_1^\dagger\prod_{n>2}(u_n+v_n\psi_n^\dagger\psi_{-n}^\dagger)|0\rangle$. As the hybridization between the bound states or the exchange coupling increases, the lower energy state becomes occupied and the ground state is $|\Psi_1\rangle$. When both states are below the chemical potential the ground state then becomes $|\Psi_{1,2}\rangle$. To determine the total energy of the system, one can diagonalize the Hamiltonian using a Bogoliubov transformation, $H=\sum_{n}\epsilon_n(\theta)\left(\gamma^\dagger_n\gamma_n-\frac{1}{2}\right)$,\cite{yaoPRL14}
where $\epsilon_n$ is the energy of state $n$. The ground state energies are therefore \cite{salkolaPRB97,yaoPRL14} 
\eq{
\mathcal{E}_{gr}(\theta)=-\frac{1}{2}\sum_n |\epsilon_n(\theta)|.
}

\section{Weak Coupling Limit}
\label{weakcouplinglimit}
To obtain the energetically favorable magnetic orientation in the weak coupling limit, we solve Eq.~(\ref{EoM2}) of the main text for the in-gap energies and expand to second order in $1/k_Fr$. We find the bound state energies are
\eq{
\epsilon_n(\theta)\approx E_n +\Delta_0( A_n +B_n \cos\theta)\left(\frac{1}{k_F r}\right)^2\,,
}
with
\begin{align}
A_1&=\frac{-2\alpha_1^2\alpha_2^2(1-\alpha_2^4)+ 2\alpha_1\alpha_2(1+\alpha_1^4-2\alpha_1^2\alpha_2^2)\cos2k_Fr}{(1+\alpha_1^2)^2[\alpha_2^2(1+\alpha_1^4)-\alpha_1^2(1+\alpha_2^4)]},\nonumber\\
B_1&=\frac{-2\alpha_2\alpha_1^3(1+\alpha_1^2-\alpha_2^2-\alpha_1^2\alpha_2^2)(1-\alpha_{2}^2)+2\alpha_2\alpha_1^3(1-\alpha_1^2+\alpha_2^2-\alpha_1^2\alpha_2^2)\cos2k_Fr}{(1+\alpha_1^2)^2[\alpha_2^2(1+\alpha_1^4)-\alpha_1^2(1+\alpha_2^4)]}\,,\nonumber\\
A_2&=\frac{2\alpha_1^2\alpha_2^2(1-\alpha_1^4)-2\alpha_1\alpha_2(1+\alpha_2^4-2\alpha_1^2\alpha_2^2)\cos2k_Fr}{(1+\alpha_2^2)^2[\alpha_2^2(1+\alpha_1^4)-\alpha_1^2(1+\alpha_2^4)]}\nonumber\\
B_2&=\frac{2\alpha_1\alpha_2^3(1-\alpha_1^2+\alpha_2^2-\alpha_1^2\alpha_2^2)(1-\alpha_{1}^2)-2\alpha_1\alpha_2^3(1+\alpha_1^2-\alpha_2^2-\alpha_1^2\alpha_2^2)\cos 2k_Fr}{(1+\alpha_2^2)^2[\alpha_2^2(1+\alpha_1^4)-\alpha_1^2(1+\alpha_2^4)]}\,.
\label{coeff}
\end{align}
We consider three cases: when the bare energies are both above the chemical potential, both below the chemical potential, or on opposite sides of the chemical potential. The total energy of the system,  according to Eq.~(\ref{totalE}) of the main text, is given by
\begin{equation}
\mathcal{E}(\theta)=\left\{
\begin{array}{cc}\displaystyle-[\epsilon_1(\theta)+\epsilon_2(\theta)]/2\,, & \epsilon_1>0,~\epsilon_2>0\\
\displaystyle[\epsilon_2(\theta)-\epsilon_1(\theta)]/2 \,, & \epsilon_1<0,~\epsilon_2>0\\
\displaystyle[\epsilon_1(\theta)+\epsilon_2(\theta)]/2 \,, &  \epsilon_1<0,~\epsilon_2<0 \end{array}\right.\,.
\end{equation}
In all cases, the total energy is extremized when $\theta=0,~\pi$, and for no intermediate values of $\theta$. To determine the energetically favored magnetic configuration, we calculate $\delta\mathcal{E}\equiv\mathcal{E}(0)-\mathcal{E}(\pi)$. When both energies are above the chemical potential,
\begin{align}
\frac{\delta\mathcal{E}}{\Delta_0}&=2\alpha_1\alpha_2\left(\frac{1}{k_Fr}\right)^2\left[\frac{1+\alpha_1^2+\alpha_2^2+2\alpha_1^2\alpha_2^2+\alpha_1^4\alpha_2^2+\alpha_1^2\alpha_1^4+\alpha_1^4\alpha_2^4}{(1+\alpha_1^2)^2(1+\alpha_2^2)^2(1-\alpha_1^2\alpha_2^2)}\right.\nonumber\\&\left.-\frac{1-\alpha_1^2-\alpha_2^2-6\alpha_1^2\alpha_2^2-\alpha_1^4\alpha_2^2-\alpha_1^2\alpha_2^4+\alpha_1^4\alpha_2^4}{(1+\alpha_1^2)^2(1+\alpha_2^2)^2(1-\alpha_1^2\alpha_2^2)}\cos2k_Fr\right]\nonumber\\
&=2\alpha_1\alpha_2\left(\frac{1}{k_Fr}\right)^2\left[\frac{1+\alpha_1^4\alpha_2^4}{(1+\alpha_1^2)^2(1+\alpha_2^2)^2(1-\alpha_1^2\alpha_2^2)}(1-\cos2k_Fr)\right.\nonumber\\
&\left.+\frac{\alpha_1^2+\alpha_2^2+2\alpha_1^2\alpha_2^2+\alpha_1^4\alpha_2^2+\alpha_1^2\alpha_2^4}{(1+\alpha_1^2)^2(1+\alpha_2^2)^2(1-\alpha_1^2\alpha_2^2)}(1+\cos2k_Fr)+\frac{4\alpha_1^2\alpha_2^2}{(1+\alpha_1^2)^2(1+\alpha_2^2)^2(1-\alpha_1^2\alpha_2^2)}\cos2k_Fr\right]\nonumber\\
&=2\alpha_1\alpha_2\left(\frac{1}{k_Fr}\right)^2\left[\frac{(1-\alpha_1^2\alpha_2^2)^2}{(1+\alpha_1^2)^2(1+\alpha_2^2)^2(1-\alpha_1^2\alpha_2^2)}(1-\cos2k_Fr)\right.\nonumber\\
&\left.+\frac{\alpha_1^2+\alpha_2^2+4\alpha_1^2\alpha_2^2+\alpha_1^4\alpha_2^2+\alpha_1^2\alpha_2^4}{(1+\alpha_1^2)^2(1+\alpha_2^2)^2(1-\alpha_1^2\alpha_2^2)}(1+\cos2k_Fr)\right]>0
\label{app_del1}
\end{align}
because $\alpha_1,~\alpha_2<1$. Analogously, when $\epsilon_1,~\epsilon_2<0$,
\begin{align}
\frac{\delta\mathcal{E}}{\Delta_0}&=-2\alpha_1\alpha_2\left(\frac{1}{k_Fr}\right)^2\left[\frac{(1-\alpha_1^2\alpha_2^2)^2}{(1+\alpha_1^2)^2(1+\alpha_2^2)^2(1-\alpha_1^2\alpha_2^2)}(1-\cos2k_Fr)\right.\nonumber\\
&\left.+\frac{\alpha_1^2+\alpha_2^2+4\alpha_1^2\alpha_2^2+\alpha_1^4\alpha_2^2+\alpha_1^2\alpha_2^4}{(1+\alpha_1^2)^2(1+\alpha_2^2)^2(1-\alpha_1^2\alpha_2^2)}(1+\cos2k_Fr)\right]>0
\label{app_del2}
\end{align}
because $\alpha_1,~\alpha_2>1$ so that the preferred magnetic orientation is antiferromagnetic when the energies are on the same side of the chemical potential. Now suppose $\epsilon_1,~-\epsilon_2>0$, then we get
\begin{align}
\frac{\delta\mathcal{E}}{\Delta_0}&=2\alpha_1\alpha_2\left(\frac{1}{k_Fr}\right)^2\left[\frac{\alpha_1^2+\alpha_2^2+2\alpha_1^2\alpha_2^2+\alpha_1^4+\alpha_2^4+\alpha_1^4\alpha_2^2+\alpha_1^2\alpha_2^4}{(1+\alpha_1^2)^2(1+\alpha_2^2)^2(\alpha_1^2-\alpha_2^2)}\right.\nonumber\\
&\left.-\frac{\alpha_1^2+\alpha_2^2+6\alpha_1^2\alpha_2^2-\alpha_1^4-\alpha_2^4+\alpha_1^4\alpha_2^2+\alpha_1^2\alpha_2^4}{(1+\alpha_1^2)^2(1+\alpha_2^2)^2(\alpha_1^2-\alpha_2^2)}\cos2k_Fr\right]\nonumber\\
&=2\alpha_1\alpha_2\left(\frac{1}{k_Fr}\right)^2\left[\frac{\alpha_1^2+\alpha_2^2+2\alpha_1^2\alpha_2^2+\alpha_1^4\alpha_2^2+\alpha_1^2\alpha_2^4}{(1+\alpha_1^2)^2(1+\alpha_2^2)^2(\alpha_1^2-\alpha_2^2)}(1-\cos2k_Fr)\right.\nonumber\\
&\left.+\frac{\alpha_1^4+\alpha_2^4}{(1+\alpha_1^2)^2(1+\alpha_2^2)^2(\alpha_1^2-\alpha_2^2)}(1+\cos2k_Fr)-\frac{4\alpha_1^2\alpha_2^2}{(1+\alpha_1^2)^2(1+\alpha_2^2)^2(\alpha_1^2-\alpha_2^2)}\cos2k_Fr\right]\nonumber\\
&=2\alpha_1\alpha_2\left(\frac{1}{k_Fr}\right)^2\left[\frac{\alpha_1^2+\alpha_2^2+4\alpha_1^2\alpha_2^2+\alpha_1^4\alpha_2^2+\alpha_1^2\alpha_2^4}{(1+\alpha_1^2)^2(1+\alpha_2^2)^2(\alpha_1^2-\alpha_2^2)}(1-\cos2k_Fr)\right.\nonumber\\
&\left.+\frac{(\alpha_1^2-\alpha_2^2)^2}{(1+\alpha_1^2)^2(1+\alpha_2^2)^2(\alpha_1^2-\alpha_2^2)}(1+\cos2k_Fr)\right]<0
\end{align}
because $\alpha_2>\alpha_1$. Therefore, making a similar argument when $\epsilon_1<0$ and $\epsilon_2>0$, when the bare energies are on opposite sides of the chemical potential and sufficiently well separated, the impurities prefer to be oriented ferromagnetically.

In the special case when $\alpha_1=\alpha_2\equiv \alpha$, the energy levels diverge according to Eq.~(\ref{coeff}). The expansion of the bound state energies is instead given by
\begin{align}
\epsilon_{n}(\theta)\approx E_n&+(-1)^n4\alpha^2\Delta_0\frac{|\cos(\theta/2)| }{1+\alpha^2}\frac{\sin k_Fr}{k_F r}\nonumber\\
&+\alpha^2\Delta_0\left[2\alpha^2\frac{1-(1-2\alpha^2)\cos2k_F r}{(1-\alpha^2)^2(1+\alpha^2)^3}-\frac{1+\alpha^4-(1-4\alpha^2+\alpha^4)\cos2k_Fr}{(1-\alpha^2)^2(1+\alpha^2)^3}\cos\theta\right]\left(\frac{1}{k_F r}\right)^2\,.
\end{align}
Although the leading order term contribution is of order $\exp(-r/\xi)/k_F r$ and oscillates with $2\pi$ periodicity in $\theta$, the difference in total energy between the parallel and antiparallel configurations, when $\alpha<1$ ($\alpha>1$), again reduces to Eq.~(\ref{app_del1}) [Eq.~(\ref{app_del2})] upon taking $\alpha_1,~\alpha_2 \rightarrow\alpha$.

\section{Strong Hybridization Expressions}
\label{strong_hyb_const}
For strongly hybridized identical impurities, the equations for the bound state energies are Eqs.~(\ref{Eferr}) and~(\ref{Eanti}), where
\begin{align}	
a&=\alpha\left\{\alpha^2\left[1+\left(\frac{1}{k_F r}\right)^2\cos 2k_F r\right]-1\right\}\nonumber\\
b&=\alpha\frac{\sin k_F r}{k_F r}\left\{\alpha^2\left[\left(\frac{1}{k_F r}\right)^2-1\right]-1\right\}\nonumber\\
c&=\alpha^2\left[2+\left(\frac{1}{k_F r}\right)^2\left(\cos 2k_F r-1\right)\right]\nonumber\\
d&=\alpha^4(1+2\cos2k_Fr)\left(\frac{1}{k_F r}\right)^4\,.
\end{align}

\section{Additional Numerics}

We plot the difference in energies between ferromagnetic and antiferromagentic configurations  $\delta\mathcal{E}$ and $\delta\mathcal{E}_{gr}$ for the same parameters as in Fig. 4 of the main text, {\it i.e.} a lattice of size $N_x\times N_y=33\times 25$ with the chemical potential $\mu/t=1$, the superconducting gap $\Delta_0/t=0.1$, see Fig.~\ref{supp1}. The exchange interaction strength is chosen to be $\bar J/t=4$. The difference between the YSR state energies in the collinear magnetic configurations is positive for nearly all values of $r$, indicating the antiferromagnetic orientation is preferred, which agrees with the analytics in the weak hybridization picture. The magnitude of the oscillations becomes larger if the quasiparticle contributions is included but the gap kept constant $\Delta_i=\Delta_0$. Upon including renormalization of the gap, $\delta \mathcal{E}_{gr}$ is significantly increased, similar to $\bar J/t=1$, again emphasizing the importance of the gap renormalization.

\begin{figure}[ht]
\centering
\begin{minipage}[b]{0.45\linewidth}
\includegraphics[width=\textwidth]{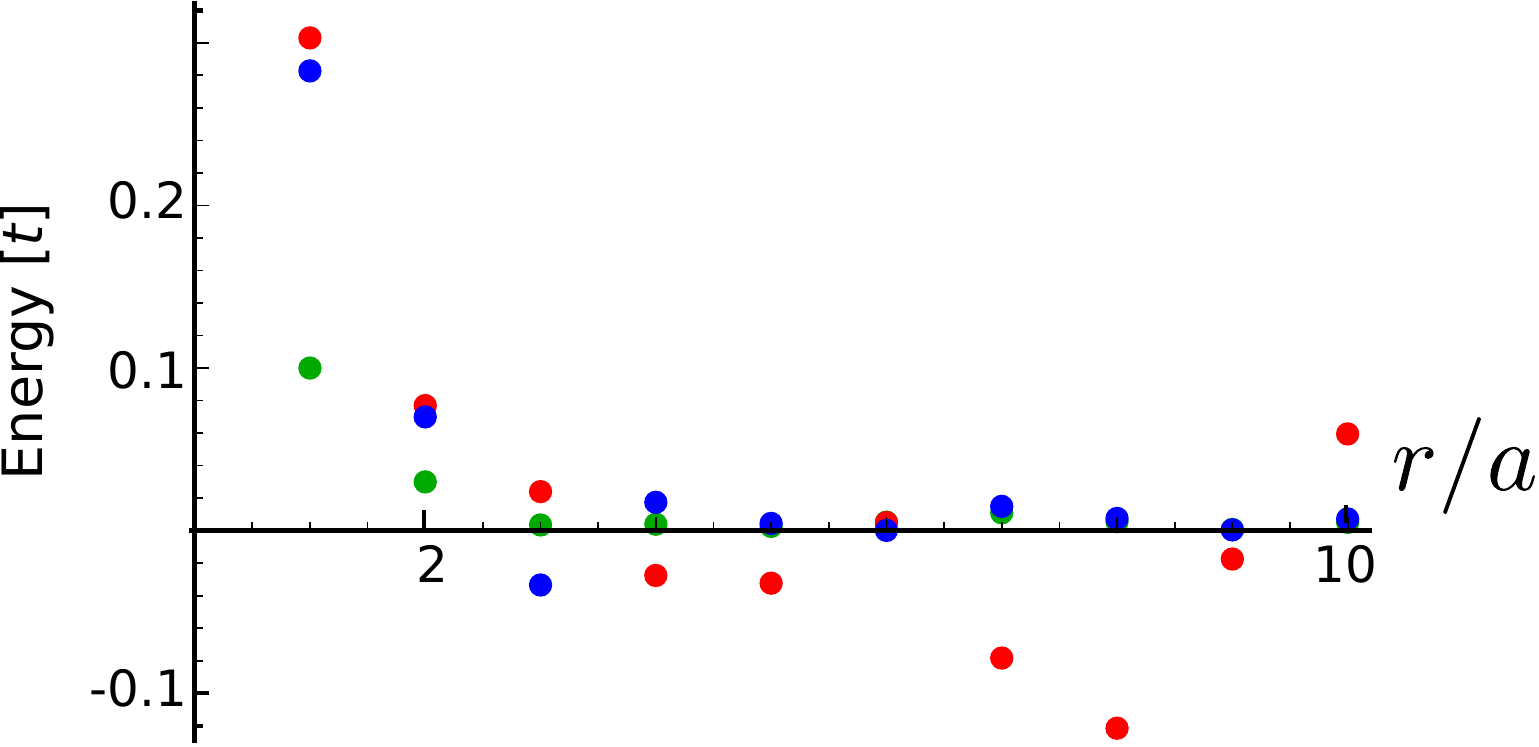}
\caption{The energy difference between ferromagnetic and antiferromagnetic configurations. The parameters are the same as in Fig. 4 of the main text with $\bar J/t=4$.}
\label{supp1}
\end{minipage}
\quad
\begin{minipage}[b]{0.35\linewidth}
\includegraphics[width=\textwidth]{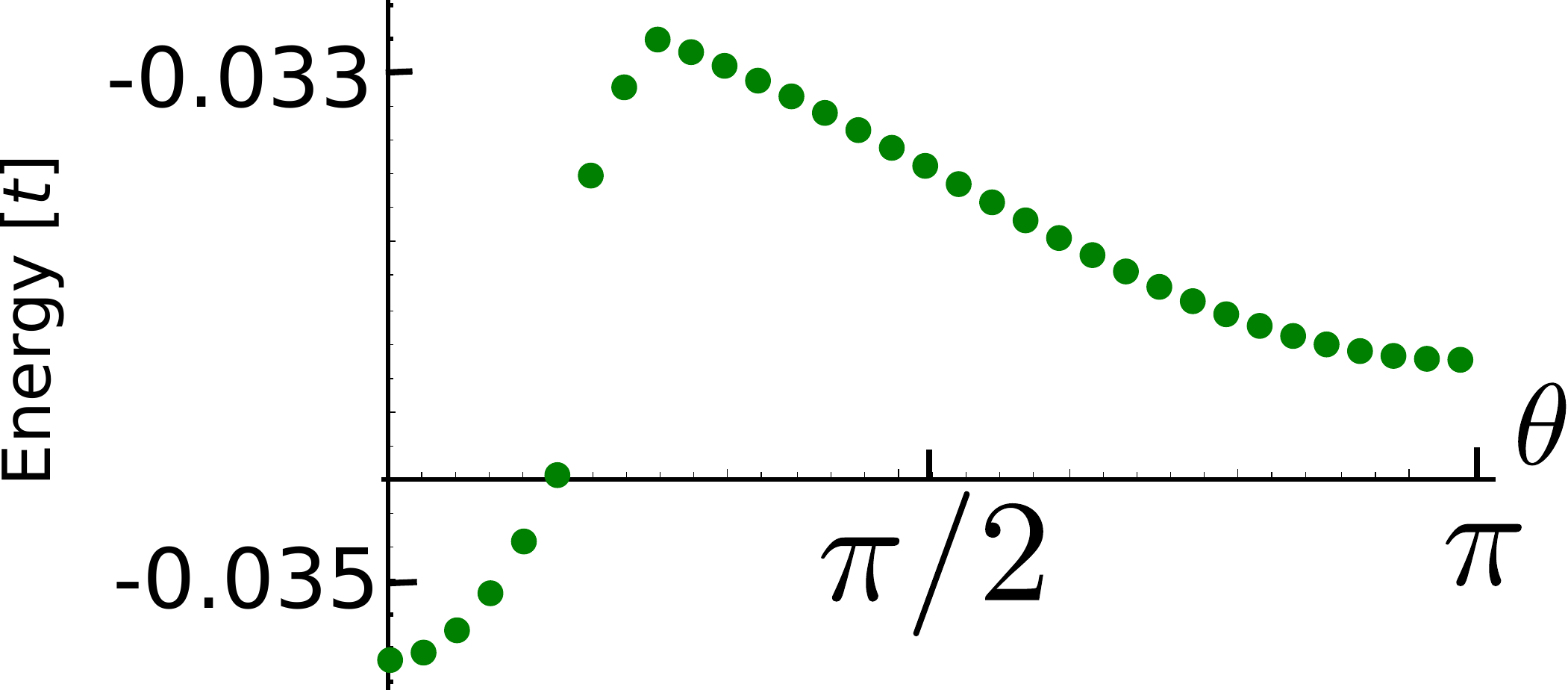}
\caption{The sum of two YSR state energy found self-consistently. The parameters are the same as in Fig. 5 of the main text with $\bar J/t=2.17$.}
\label{supp2}
\end{minipage}
\end{figure}

Taking $\bar J/t=2.7$ while leaving all other parameters the same as in Fig.~5 of the main text, we plot $\mathcal{E}$ as a function of $\theta$. Similar to the analytics, we find a ground state and a metastable state at the collinear configurations of the magnetizations. However, because of the self-consistent renormalization of the gap, there is a jump in $\mathcal{E}$ at $\theta\approx\pi/6$ where the QPT occurs. We note that the change in $\mathcal{E}$ is several orders of magnitude smaller as compared with $\mathcal{E}_{gr}$, see Fig.~5 of the main text.
\end{widetext}
\end{document}